\documentclass[epj-spec]{svjour}
\usepackage{graphicx}
\usepackage{axodraw4j}
\usepackage{wrapfig}
\usepackage{subfig}

\newcommand{\Kshort}{\ensuremath{K^0_S}}
\newcommand{\Klong}{\ensuremath{K^0_L}}

\newcommand{\MeV}{\ensuremath{\textrm{MeV}}}

\newcommand{\MeVcc}{\ensuremath{\textrm{MeV}/\textrm{c}^2}}

\newcommand{\GeV}{\ensuremath{\textrm{GeV}}}
\newcommand{\GeVc}{\ensuremath{\textrm{GeV}/\textrm{c}}}
\newcommand{\GeVcc}{\ensuremath{\textrm{GeV}/\textrm{c}^2}}

\begin{document}
\title{The COMPASS sandwich veto detector and a first look at kaonic
  final states from a $\pi^-$(190 GeV) beam on a
  proton target}
\author{Tobias Schl\"uter\thanks{\email{tobias.schlueter@physik.uni-muenchen.de}} for the COMPASS collaboration}
\institute{LMU Munich}
\abstract{
We introduce the sandwich veto detector that was built for the 2008
and 2009 hadron runs of the COMPASS experiment at CERN.  During these
beamtimes it was serving as a veto detector for neutral and charged
particles outside the spectrometer acceptance, mostly thought to
originate from reactions which excited the target.  We also
present first mass spectra from $\pi^-(190\,\GeV) p \to \pi^- 
\Kshort \Kshort p$ that were measured in the 2008 hadron run.}
\maketitle
\section{Introduction}
\label{intro}
The COMPASS experiment at CERN is a multi-purpose, two-stage
spectrometer pursuing a variety of fixed-target physics
programmes~\cite{Abbon:2007pq}.  Its 2008 and 2009 Pion runs were
dedicated to studying mesonic states produced both as diffractive
excitations of the scattered beam pion and as objects produced
centrally between the scattered beam pion and the recoiling proton
target.  In order to suppress events with particles lying outside the
spectrometer acceptance, the sandwich veto detector was installed
between the target recoil detector~\cite{Bernhard:2009} and the
spectrometer.  The design of the detector and its performance in the
2008 pion run are discussed in sec.~\ref{sec:sandw-veto-detect}.  As
an illustration of both the performance of the COMPASS spectrometer as
a whole and the current status of our analyses, we present some
preliminary results concerning neutral kaonic final states in
sec.~\ref{sec:first-look-at} and discuss the prospects for partial
wave analyses of those states.

\section{The Sandwich Veto Detector}
\label{sec:sandw-veto-detect}

The acceptance of the COMPASS spectrometer as seen from the target is
approximately conical with the beam direction along the symmetry axis
of the cone and an opening angle of the cone of 11 degrees as seen
from the center of the liquid hydrogen target.  The recoil proton
detector (RPD) covers angles above 55 degrees.  In order to reject events
where a particle ends up between these two acceptances, a veto
detector had to be placed between the RPD and the spectrometer.  Since
it had to detect both charged (mostly pions) and neutral particles
(mostly photons from neutral pion decays), a thin electromagnetic
calorimeter was designed for this purpose, the sandwich veto detector.
The design criteria differed from an electromagnetic calorimeter as
found e.g.\ in both of the stages of the COMPASS spectrometer insofar,
as energy resolution (and therefore good shower containment, high
radiation length) could be sacrificed in order to allow for fast
readout and the little space available in the direction along the beam
direction.  Likewise, spatial resolution was not a requirement in the
design.  On the other hand, high detection efficiency for
low-energetic photons was required, extending below $100\,\MeV$
together with high rate capability.

\begin{figure}[tbp]
  \centering
  \subfloat[Schematical cross section of a module.  Lead is depicted in
  dark gray, steel in light gray, scintillators in blue.  Target to
  the left, spectrometer to the right.  The step structure near the bottom
  indicates the shape chosen to match the spectrometer acceptance.]  {
    \includegraphics[width=.25\textwidth]{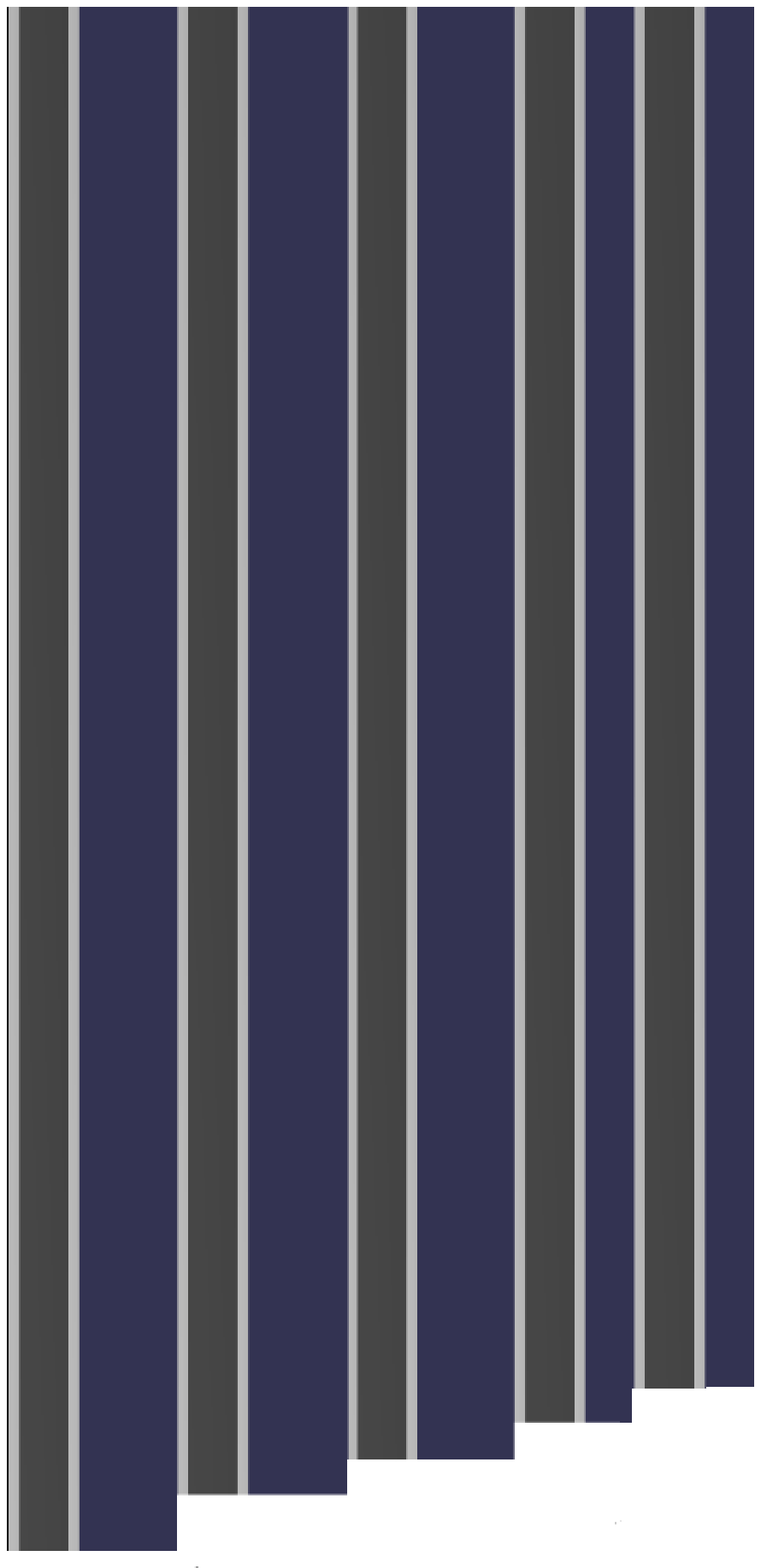}
    \label{fig:cross-section}
  } \quad \subfloat[MIP response.  The picture shows the integrated
  response of one PMT as recorded by the SADC readout for events where
  a muon hit the corresponding detector module.  The axes are labelled
  in arbitrary units.  The distribution is broadened from the expected
  Landau distribution due to statistical fluctuations of the light
  collection.  The clear separation from zero is indicative of the
  high efficiency.]  {
    \includegraphics[width=.68\textwidth]{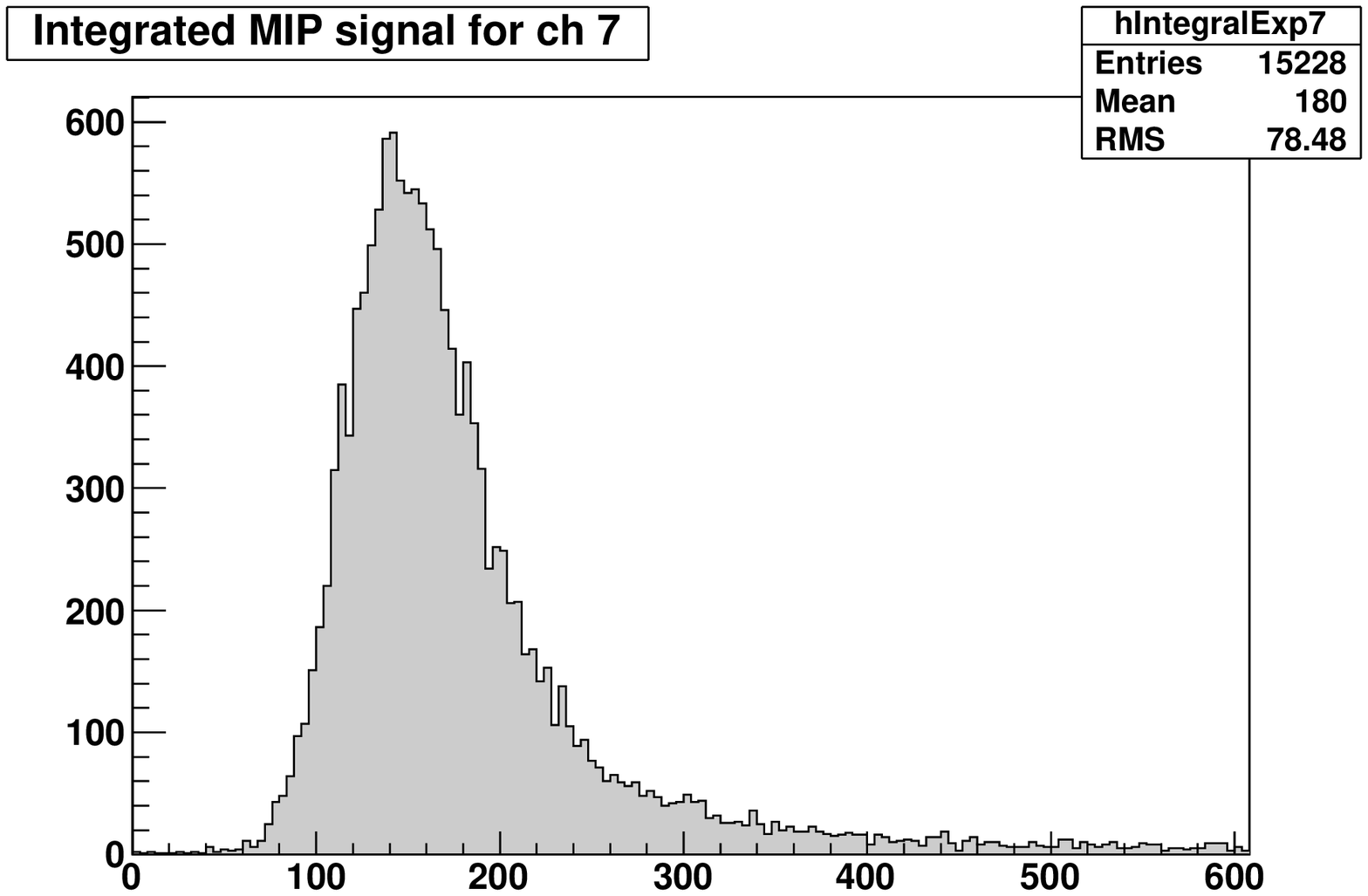}
    \label{fig:landau}
  }
  \caption{Design and performance of the sandwich veto detector.}
\end{figure}
In order to achieve these goals, the following design was settled
upon: the $2\times 2\,\textrm{m}^2$ active surface of the detector was
divided into 12 modules, each read out by 208 optically active
wavelength shifting optical fibres converging on a single
photomultipier.  A central hole, corresponding to the spectrometer
acceptance, was left empty.  Each module consists of a
sandwich of five lead layers alternating with plastic scintillators.
Each lead layer has steel plates glued to it on both sides in order to
ensure mechanical stability.  The design is illustrated in
fig.~\ref{fig:cross-section}.  In order to economize in thickness, the
two most downstream scintillator layers are implemented at half
thickness, compared to the other layers' thickness of
$10\,\textrm{mm}$.  This choice was made after confirming in Monte
Carlo that this combination gives the best efficiency for low-energy
photons.  The lead layers are $5\,\textrm{mm}$ each, the steel layers
$1\,\textrm{mm}$.  The optical fibres are glued into grooves cut into
the scintillators.

With this setup we estimated nearly perfect detection efficiency for
minimum ionizing particles (MIPs), while being only slightly
susceptible to false vetoes from $\delta$-electron irradiation.  These
predictions were confirmed in the course of the experiment.  COMPASS
uses a wide high-energy muon beam for alignment that illuminates most
of the surface of the sandwich detector.  Using this beam, we
confirmed a MIP detection efficiency of $> 98\%$ after accounting for
inefficiencies along the edges of the individual detector modules by
comparing hits predicted from COMPASS's tracking with actual hits
recorded in the sandwich veto detector.  This degree of efficiency,
together with the electronic thresholds used for the veto signal lead
-- by comparison with our Monte Carlo studies of energy deposits -- to
an efficiency of $> 70\%$ for 100\,\MeV{} photons.  Photon energy
deposits increase steeply with increasing photon energy, as does the
detection efficiency.

Comparing the trigger rates for various triggers in the COMPASS
experiment with and without the sandwich veto detector included in the
triggers leads to the following figures of merit:
\begin{enumerate}
\item enrichment factor 3.2: the trigger rates for the main physics
  trigger DT0 increase by a factor 3.2 if the sandwich is not included
  in the veto.
\item false vetoes happen for about 2\% of all events.  This lies
  within Monte Carlo expectations for rates induced by
  $\delta$-electrons, and has been verified in different physics
  scenarios.
\item the performance of the detector doesn't seem to suffer from the
  increased rates during the 2009 run.
\end{enumerate}

Ongoing studies of the sandwich veto concern its light collection
efficiency and a detailed understanding of its geometrical acceptance.
After the 2009 run of the experiment it was removed from the
experimental area as it is not needed for the COMPASS muon run that is
going to take place in 2010.

\section{A first look at kaonic final states}
\label{sec:first-look-at}

\subsection{Motivation}
\label{sec:motivation}

Data collected in the 190\,\GeVc{} $\pi^-p$ run of 2008 allow high
statistics studies of kaonic channels in diffractive and central
processes.  It is useful to remember the allowed quantum number
combinations of neutral $K \bar K$ systems (all with isospin $I = 0$
or $1$):
\begin{center}
\begin{tabular}[c]{r|cccccc}
  state & \multicolumn{5}{c}{allowed $J^{PC}$} \\
\noalign{\smallskip}
 \hline\noalign{\smallskip}
  $\Kshort \Kshort$ & $0^{++}$ &        & $2^{++}$ &        & $4^{++}$\\
  $\Kshort \Klong$  &        & $1^{--}$ &        & $3^{--}$ & \\
  $K^+ K^-$         & $0^{++}$ & $1^{--}$ & $2^{++}$ &
  $3^{--}$&  $4^{++}$ \\
 \hline
\end{tabular}
\end{center}

Previous work on the centrally produced $K^+K^-$ system at
$450\,\GeVc$ by WA102 resulted in a $q \bar q$ ($ q = u,d,s $) and
glueball mixing scheme for the $0^{++}$ states observed at 980, 1370,
1500, and
1710\,\MeVcc~\cite{Barberis:1999cq,Barberis:1999am,Close:2001ga}.
Open questions concern the resonance nature of $f_0(1370)$, with
inconsistent decay branchings in other production
processes~\cite{Bugg:2007ja,Klempt:2007cp}, and the extension of the
level scheme to the 2\,\GeVcc{} region, in particular for the $0^{++}$
and $2^{++}$ states which are the quantum numbers of the lightest
expected glueballs.

Preferential $K \bar K$ decay of glueballs was recently predicted on
the basis of chirality arguments, in contrast to earlier expectations
of flavour blind decay~\cite{Chanowitz:2005du}.  The \Kshort \Kshort{}
(\Kshort \Klong{}) system is particularly suited for spectroscopy
because of its selectivity for $C = +\ (-)$ states.  From previous
central production studies, only poor statistics data exist on \Kshort
\Kshort{}. COMPASS data outdo these by more than an order of magnitude
and are virtually free of background, but at the price of no clear
separation between diffractively and centrally produced final states.

The $K \bar K \pi$ system, as produced in diffractive $\pi^- p$
collisions, can exist with exotic quantum numbers $1^{-+}$ .
Preferential $K^*(\to K \pi) K$ decay is expected from models of
quartet and hybrid states~\cite{Chung:2002fz,Close:1994hc}.  So far no
kaonic exotics emerged from various experimentally studied production
processes~\cite{Klempt:2007cp}.  Since their existence is
inevitable in flavour multiplet schemes, they can be considered as
touchstones of quartet models.  COMPASS produced $\Kshort\Kshort\pi^-$
with high statistics, well suited for partial wave analysis.  In
addition to exotics, states or dublets with $q \bar q$ quantum numbers
like the enigmatic $E/i(1405)$ or $\pi(1800)$ will possibly make an
appearance in these data.

In order to disentangle the angular distributions of the final states,
and thereby the physical content, partial wave analysis needs to be
performed.  For this we will make use of two software packages that
are currently being developed within our collaboration.

At this conference we give a first impression of the status of our
analysis of the $\Kshort\Kshort$ channel, but analyses of the sibling
$K^+K^-$ and $\Klong\Kshort$ channels are also underway.  The
$\Kshort\Kshort$ channel is the easiest of these in terms of
selection, as the two displaced vertices of the preferred \Kshort{}
decay into a pair of charged pions are a fairly unique signature.

\subsection{Event selection}

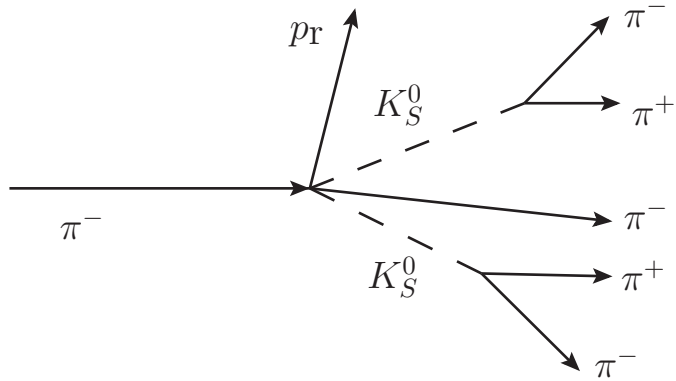
\begin{figure}
  \centering
  \begin{picture}(271,151) (15,2)
     \SetWidth{1.0}
     \SetColor{Black}
     \Line[arrow,arrowpos=0.965,arrowlength=5,arrowwidth=2,arrowinset=0.2](16,71)(128,71)
     \Line[arrow,arrowpos=1,arrowlength=5,arrowwidth=2,arrowinset=0.2](128,71)(144,135)
     \Line[dash,dashsize=10](128,71)(208,103)
     \Line[arrow,arrowpos=1,arrowlength=5,arrowwidth=2,arrowinset=0.2](128,71)(237,59)
     \Line[dash,dashsize=10](128,71)(192,39)
     \Line[arrow,arrowpos=1,arrowlength=5,arrowwidth=2,arrowinset=0.2](208,103)(237,133)
     \Line[arrow,arrowpos=1,arrowlength=5,arrowwidth=2,arrowinset=0.2](208,103)(240,103)
     \Line[arrow,arrowpos=1,arrowlength=5,arrowwidth=2,arrowinset=0.2](192,39)(237,38)
     \Line[arrow,arrowpos=1,arrowlength=5,arrowwidth=2,arrowinset=0.2](192,39)(226,5)
     \Text(35,52)[lb]{\Large{\Black{$\pi^-$}}}
     \Text(121,125)[lb]{\Large{\Black{$p_\textrm{r}$}}}
     \Text(152,95)[lb]{\Large{\Black{\Kshort}}}
     \Text(150,30)[lb]{\Large{\Black{\Kshort}}}
     \Text(246,132)[lb]{\Large{\Black{$\pi^-$}}}
     \Text(249,95)[lb]{\Large{\Black{$\pi^+$}}}
     \Text(246,56)[lb]{\Large{\Black{$\pi^-$}}}
     \Text(245,33)[lb]{\Large{\Black{$\pi^+$}}}
     \Text(235,0)[lb]{\Large{\Black{$\pi^-$}}}
  \end{picture}

  \caption{Selected event topology}
  \label{fig:topology}
\end{figure}

\begin{wrapfigure}{r}{0.6\textwidth}
  \centering
  \includegraphics[width=0.55\textwidth]{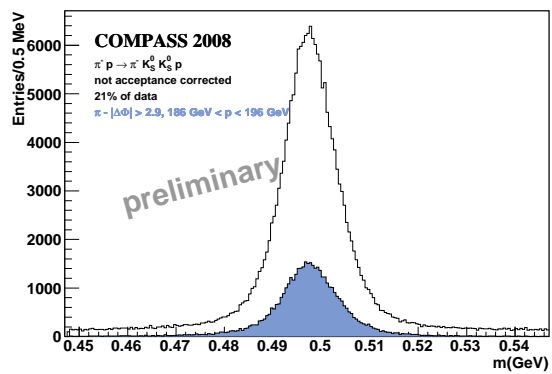}
  \caption{Distribution of reconstructed \Kshort{} masses.  The upper
    histograms shows the distribution before requiring exclusivity as
    explained in the text, the filled histogram shows the final sample.}
  \label{fig:mKshort_comparison}
\end{wrapfigure}

The analysis so far is based on two weeks of data taken during the
2008 negative hadron beam run of the COMPASS experiment, which amounts
to some 21\% of the 2008 data collected with negative pion beam.
These data together with 2009's will allow high-statistics
partial-wave analyses.  From the reconstructed data we selected events
with a topology as depicted in fig.~\ref{fig:topology}, i.e.\ one
vertex with an incoming beam track and a single outgoing track, and
two displaced $V^0$ vertices.  We have not included \Kshort{}s that
decayed so close to the primary vertex that the reconstruction
software could not separate their decay from the original interaction.
This mostly affects the acceptance of slow \Kshort{}s that decay close
to their point of creation and we are planning to include such events
should acceptance of slow \Kshort{}s turn out to be an issue.  We
ensured that the \Kshort{}s' point of origin was the primary vertex by
requiring that $\cos\theta > 0.9995$ for the angle $\theta$ between
the reconstructed momentum of each \Kshort{} and the line connecting
its decay vertex to the primary vertex.  In a first selection step all
events with exactly two such \Kshort{} candidates with reconstructed
\Kshort{} mass within 5\,\MeVcc{} of the PDG value were selected.

In order to further suppress pollution from the small Kaon admixture
($O(3\%)$) in the beam we required the beam-line CEDAR
detectors~\cite{Jasinski:2009} to not have identified the beam
particle as Kaon.  Exclusivity was ensured by requiring the total
momentum of the final state particles to be within $185\,\GeV$ and
$195\,\GeV$ (nominal beam momentum $191\,\GeV$).  Additionally, we
used the recoil proton detector to verify the planarity of the event,
i.e.\ we required that the direction of the recoil proton lay in the
plane formed by the beam and the total momentum of the pions (for the
azimuthal angle $\Delta\Phi$ between the plane of the recoil track and
that of the spectrometer system: $\pi - |\Delta\Phi| >
2.9\,\textrm{rad}$).  A total of 22\,914 events passed this selection
for the reduced data set used for these preliminary studies.  This
already exceeds the statistics gathered by the WA102
collaboration~\cite{Barberis:1999am} for their sample of centrally
produced $\Kshort\Kshort$ by approximately an order of
magnitude.

As an example of the degree of purity of the selection, we show the
reconstructed masses of the selected \Kshort{}s in
fig.~\ref{fig:mKshort_comparison} which shows that the final selection
is both virtually free of background with a reconstructed mass nicely
centered at the PDG value~\cite{Amsler:2008zzb}.

\subsection{Kinematics of the final state}
\label{sec:results}

\begin{figure}[tbp]
  \centering

  \subfloat[Rapidity distribution.  NB: for each event there are two
  entries corresponding to \Kshort{}s.]  {
    \includegraphics[width=0.45\textwidth]{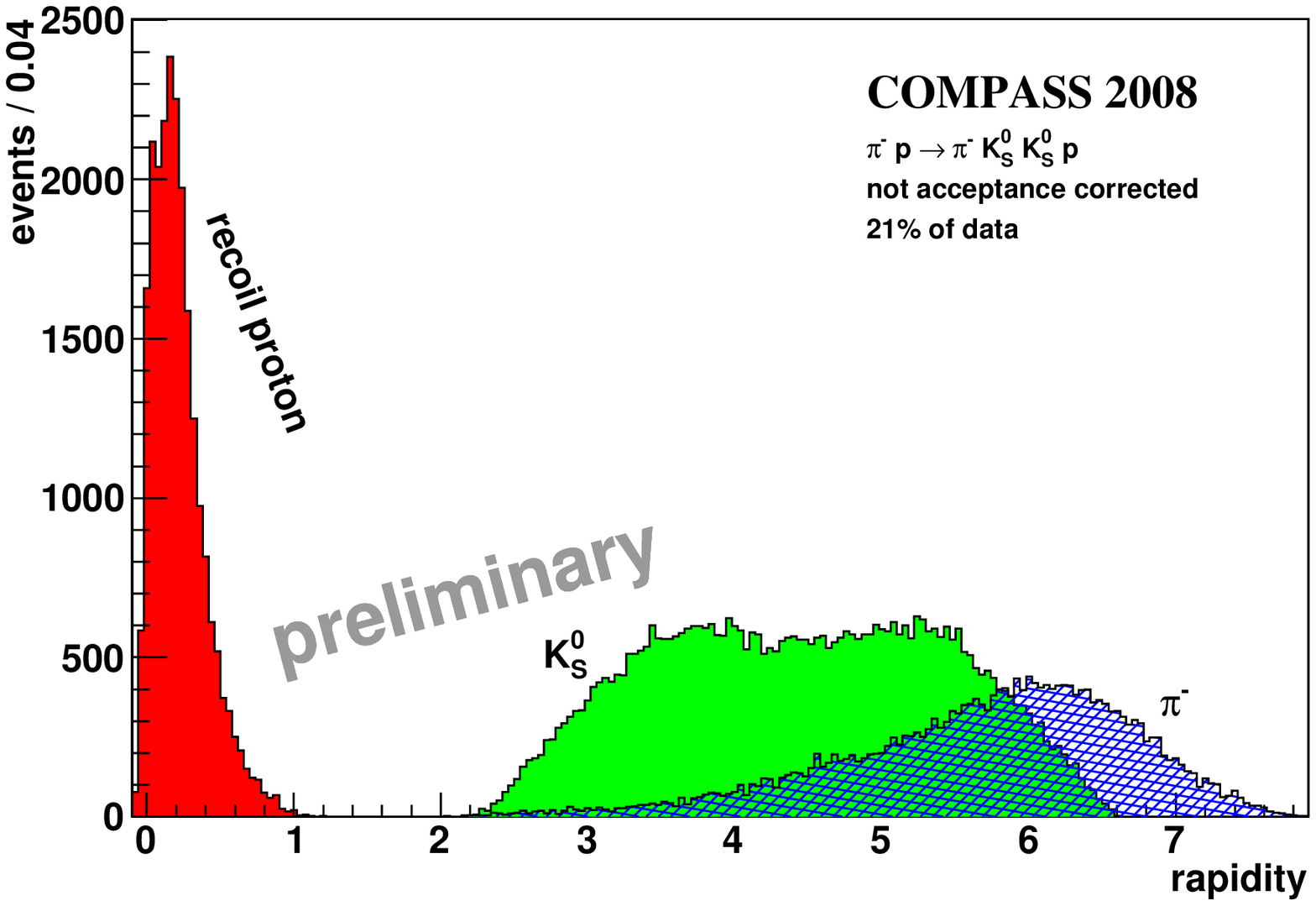}
    \label{fig:rap_all.eps}
  }
  \quad
  \subfloat[Invariant mass sepctrum of the $KK$ subsystem.]
  {
    \includegraphics[width=0.45\textwidth]{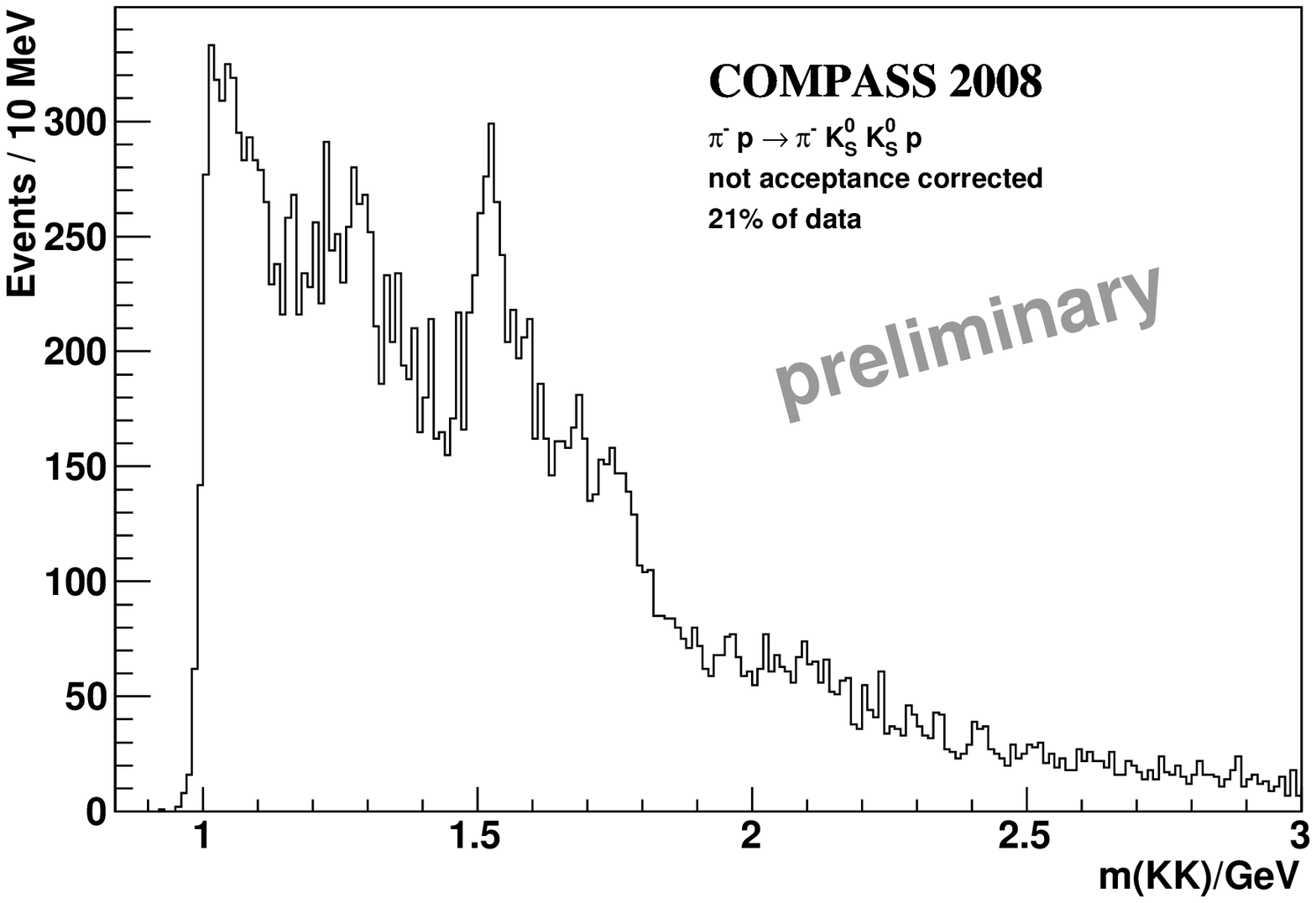}
    \label{fig:mKK_all}
  }
  \caption{Distributions illustrating the properties of the overall
    sample.  Refer to text for selection.}
\end{figure}

The distribution of rapidities of the particles in the selected events
is depicted in fig.\ \ref{fig:rap_all.eps}.  It shows a clear
separation between the rapidity of the recoil proton and the
spectrometer system.  On the other hand, the rapidities of the forward
particles shows no clear separation of different regimes.  Overall,
the pions have higher rapidity than the \Kshort{}s, but there's a
large fraction of overlap.  On the other hand the corresponding $KK$
mass spectrum reproduces well-known resonances fairly faithfully as
show in fig.\ \ref{fig:mKK_all}.

\begin{figure}[htbp]
  \centering
  \includegraphics[width=0.7\textwidth]{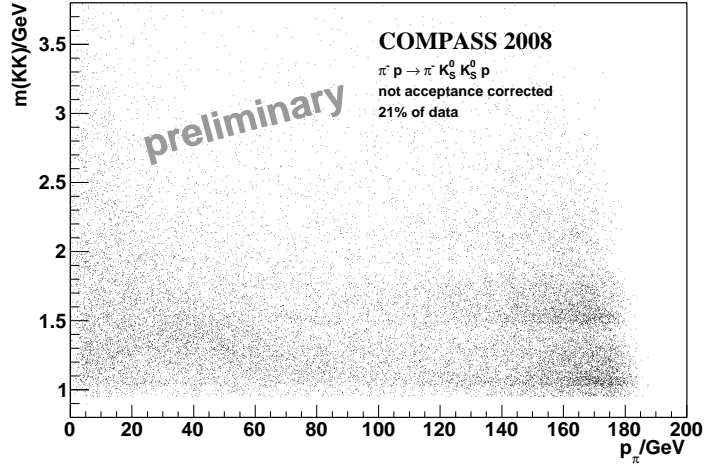}
  \caption{Invariant mass of the $K\bar K$ system against the momentum of the final state $\pi^-$.  The presence of different regimes is striking.}
  \label{fig:mKK_vs_ppi}
\end{figure}

\begin{figure}[tbp]
  \centering

  \subfloat[Rapidity distribution.  NB: for each event there are two
  entries corresponding to the \Kshort{}s.]  {
    \includegraphics[width=0.45\textwidth]{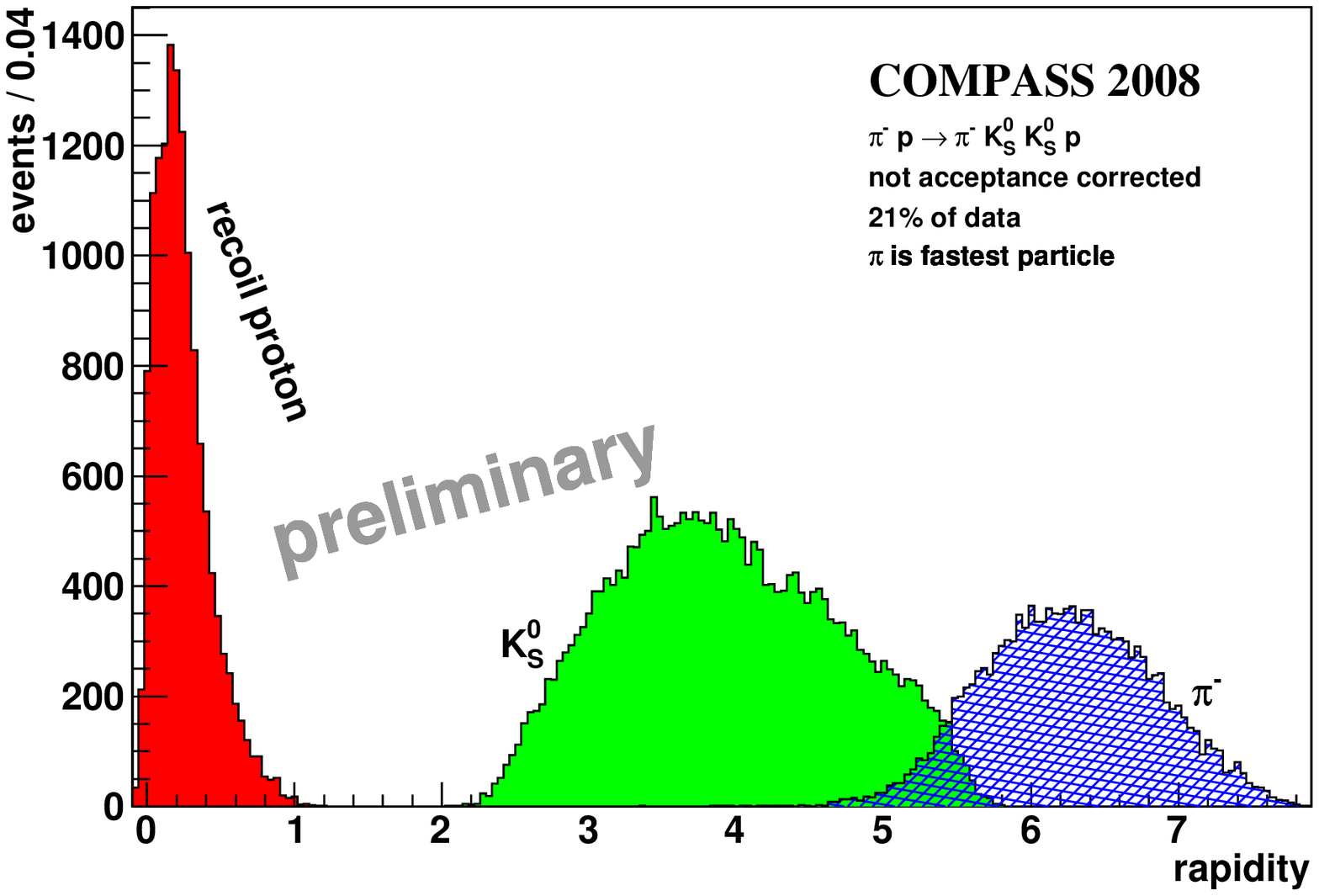}
    \label{fig:rap_pifastest}
  }
  \quad
  \subfloat[Invariant mass sepctrum of the $\Kshort\Kshort$ subsystem.]
  {
    \includegraphics[width=0.45\textwidth]{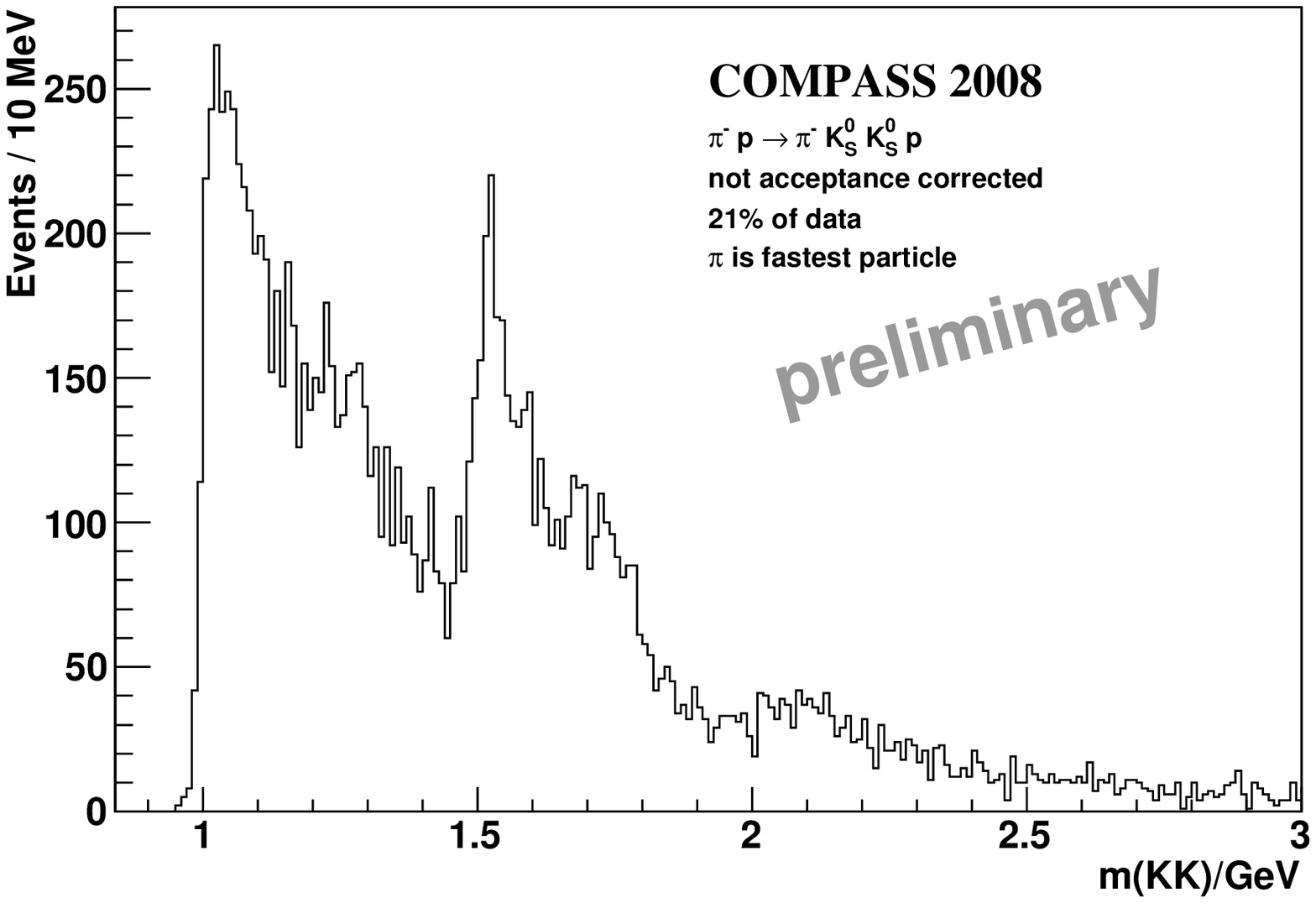}
    \label{fig:mKK_pifastest}
  }
  \caption{Distributions illustrating the properties of the sub-sample
    where the pion has the highest momentum.}
\end{figure}

The presence of different production processes comes apparent in
fig.~\ref{fig:mKK_vs_ppi} which plots the mass of the $K\bar K$ system
over the pion momentum.  The distribution looks significantly
different in the regime with low and high pion momenta.  In order to
suppress diffractive production where a three-body system
approximately carrying the beam momentum would decay into
$\Kshort\Kshort\pi^-$ we selected a subset where the final-state pion
carriest the largest momentum.  Selections using rapidity gaps were
also tried out, but the resulting mass spectra are comparable.  The
corresponding rapidity distribution and mass spectrum are depicted in
fig.~\ref{fig:rap_pifastest} and fig.~\ref{fig:mKK_pifastest}.

\section{Acknowledgements}
This research was supported by the DFG cluster of excellence 'Origin
and Structure of the Universe' (www.universe-cluster.de).
%



\end{document}